\documentstyle[aps,twocolumn,epsf]{revtex}
\newcommand{\lapx}{\,\raisebox{-.5ex}
 {$\stackrel{\raisebox{-.5pt}{$\textstyle <$}}{\sim}$}\,}
\newcommand{\gapx}{\,\raisebox{-.5ex}
 {$\stackrel{\raisebox{-1pt}{$\textstyle >$}}{\sim}$}\,}

\newcommand{\wc}{$\omega_{\rm c}$}
\newcommand{\cm}{${\rm cm}^{-1}$}

\newcommand{\K}[2]{$K_{#1#2}$}
\renewcommand{\L}[2]{$\lambda_{#1#2}$}

\newcommand{\E}[1]{$E_{#1}$}
\renewcommand{\P}[1]{$P_{#1}$}
\def\break#1{\pagebreak \vspace*{#1}}
\def\epsfig#1#2#3#4
         {
         \epsfysize=#2 \vbox{ \hglue#3 \epsfbox[#4]{#1} }
         }
\def\epsfigrot#1#2#3#4
         {
         \epsfxsize=#2
         \setbox\rotbox=\hbox to #2{\epsfbox[#4]{#1}}
         \vbox{\hglue#3 \rotl\rotbox}
         }
\newbox\rotbox
\input rotate

\begin{document}
\draft

\title{ On the Mechanism of the Primary Charge Separation in
 Bacterial Photosynthesis }

\author{C.H. Mak}
\address{ Department of Chemistry, University of Southern California,
  Los Angeles, California 90089-0482, USA }

\author{Reinhold Egger}
\address{ Fakult\"at f\"ur Physik, Universit\"at Freiburg,
 Hermann-Herder-Str.~3, D-79104 Freiburg, Germany }

\date{Date: \today}
\maketitle

\widetext
\begin{abstract}
We present a detailed analysis of the mechanism of the primary charge
separation process in bacterial photosynthesis using real-time
path integrals.  Direct computer simulations as well as an
approximate analytical theory have been employed to map out the
dynamics of the charge separation process in many regions of the
parameter space relevant to bacterial photosynthesis.  Two distinct
parameter regions, one characteristic of sequential transfer and the
other characteristic of superexchange, have been found to yield charge
separation dynamics in agreement with experiments.
Nonadiabatic theory provides accurate rate estimates
for low-lying and very high-lying bacteriochlorophyll state
energies, but it breaks down in between these two regimes.

\end{abstract}
\pacs{}
\narrowtext

\section{Introduction}

The problem of bacterial photosynthesis has attracted a lot of
recent interest since the structures of the photosynthetic reaction
center (RC) in purple bacteria {\em Rhodopseudomonas viridis} and
{\em Rhodobacterias sphaeroides} have been determined\cite{xray1}.
Much research effort is now focused on understanding the relationship
between the function of the RC and its structure.  One fundamental
theoretical question concerns the actual mechanism of the primary
electron transfer (ET)
 process in the RC, and two possible mechanisms (or a
combination of both) have emerged out of the recent theoretical work
\cite{marcus87,bixon87,bixon91,coherent,mukamel,bialek,nica}.
The first is an incoherent two-step mechanism where the charge
separation involves a sequential transfer from
the excited special pair (P$^*$) via an intermediate
bacteriochlorophyll monomer (B)
to the bacteriopheophytin (H).
The other is a coherent one-step superexchange mechanism,
with P$^+$B$^-$ acting only as a virtual intermediate.

{}From experimental studies to date, one cannot easily rule out
either possibility.  In fact, conflicting interpretations have been
derived from different experiments.  The detection of transient
population in the P$^+$B$^-$ state has been considered by many to be
the key experimental evidence that can differentiate between the two
mechanisms.  Many transient absorption spectroscopic experiments
\cite{breton,kirmaier} could not detect bleaching in the B$^-$
band, lending support to the coherent superexchange mechanism.
On the other hand, some other experiments \cite{holzapfel,dressler}
point toward a two-step process  since
a fast second rate constant
corresponding to the P$^+$B$^-$H $\to$ P$^+$BH$^-$ transition
was detected.  The picture emerging from a comparison
of the molecular dynamics (MD) simulations for the RC
carried out so far \cite{chandler,warshel,schulten}
is equally murky. Different groups
arrive at conflicting conclusions.

\break{1.80in}

Despite the emphasis placed on the experimental detection of
population in the P$^+$B$^-$ band,
we have shown in a recent real-time path integral simulation\cite{JPC}
that observation of transient population in P$^+$B$^-$ cannot
definitively rule out either mechanism.  In fact, significant
population in the P$^+$B$^-$ state (up to 20\%) can be observed in
both the sequential and the superexchange regimes in the simulations.
This suggests that transient population on the P$^+$B$^-$ state, even
if detectable, may not really be useful for differentiating between
the two mechanisms\cite{bialek}.
In fact, almost all experimentally observed dynamical
features of the RC can be reproduced nicely in either regime.

In this Letter, we report a detailed path integral analysis
of the mechanism of the primary charge separation process.
We have computed the  dynamics of the primary ET
(i.e., the time-dependence of the occupation probabilities on the
three relevant chromophores) as a function of the
P$^+$B$^-$ energy for many
combinations of the electronic coupling and the reorganization energy.
The primary objective is to determine the parameter values
that are required for the dynamics to be consistent
with characteristic experimental observations, and hence to provide a
map of the various possible parameter regions in which
the RC could operate.
Our previous simulations have been carried out for a relatively small
energy range for the P$^+$B$^-$ state (between $-666$~\cm\, and $+666$~\cm\,
relative to the P$^*$ state).  In this Letter, we present new results for
a much larger range of energies, up to 3000~\cm\, above and 1333~\cm\,
below the P$^*$ state.  This spans the region usually considered to be
relevant for the RC, except for the region with extremely
high P$^+$B$^-$ energy (a value of $+8000$~\cm\,  has been suggested
in Ref.\cite{chandler}).

\section{Three-state spin-boson model for the RC}

The model for our studies has been  discussed previously in great detail
\cite{nica,chandler,JPC}. For the sake of completeness,
we will summarize some of the relevant features here.
For a model of the RC, we consider Hamiltonians of the form
\begin{equation}\label{spinbos}
H = H_0 + H_B + H_I \;,
\end{equation}
where $H_0$ describes the bare three-state system,
$H_B$ corresponds to a Gaussian bath describing the
protein environment, and
$H_I$ is a bilinear system-environment coupling describing
the interaction of the electric dipole moment with the bath
polarization.
The states 1, 2, and 3 correspond to the electronic configurations
P$^*$BH, P$^+$B$^-$H and P$^+$BH$^-$,
respectively, and we allow for arbitrary
binding energies of these states.  The bare three-state Hamiltonian is
\begin{equation} \label{freeham}
H_0 =  \left( \begin{array}{ccc}
 E_1    & -K_{12} & -K_{13} \\
-K_{12} &  E_{2}  & -K_{23} \\
-K_{13} & -K_{23} &  E_{3}
\end{array} \right) \;,
\end{equation}
where the electronic coupling (tunnel matrix element)
between states $i$ and $j$ is denoted by $K_{ij}$.

This spin-boson model is an idealized model
for the RC.  Its limitations have been discussed in Ref.\cite{JPC}.
The most significant aspect of this model, however, is its ability
to capture the effects of the low-frequency protein modes
on the tunneling electron from first principles\cite{leggett}.
In this sense, the protein casing
in the RC presents itself to the electron as a Gaussian
fluctuating polarization field
 with a continuous spectral density.  Previous MD
simulations  have shown that this spectral density is smooth,
featureless, and of Ohmic form
with a characteristic frequency \wc\, of about 160~\cm. They
 have also confirmed the assumption of Gaussian statistics
for the bath polarization since the free energy curves were always found
to be strictly parabolic.
 In the classical (high temperature) limit, the spectral density has a simple
relationship with the familiar reorganization energy \L13
between states 1 and 3, and one can characterize the Ohmic spectral
density completely by the two parameters \wc\, and \L13.
We fix a number of parameters in our model according to reasonable
experimental and theoretical estimates\cite{JPC}:
\begin{enumerate}
\item
With the energy scale fixed by setting \E1 = 0, the energy of the
P$^+$H$^-$ state \E3 is set at $-2000$~\cm.
\item
The bath frequency \wc\, is set at 166~\cm\, according to MD results.
\item
The ratio \K23/\K12 is set at 4, and \K13 is assumed to be
 negligible.  All electronic
couplings are assumed to be independent of the actual protein
configuration (Condon approximation).

\end{enumerate}

The technical details of our real-time quantum Monte Carlo (QMC)
simulations have been reported in great detail elsewhere \cite{JPC}
and will not be repeated here. This numerical technique allows
for a computation of the time-dependent occupation probabilities
on the electronic states up to a certain time limit.
QMC simulations cannot be carried out to very long times
due to the well-known dynamical sign problem.
However, the simulations reported here have been carried out up to a few
picoseconds, the same timescale as the experimentally
determined ET rate.
In these simulations, we search for parameter regions in which
the charge separation proceeds in qualitative agreement with
experimental observations.  We consider the following to be key
experimental characteristics:
(1) the P$^+$B$^-$ population (\P2) should be small throughout
(\lapx 20\%); (2) the charge separation rate should be about
3~ps at room temperature; and (3) the rate should increase about
twofold at cryogenic temperatures.
For each value of the P$^+$B$^-$ state energy \E2,
we attempt to find a combination
of \K12 and \L13 that would yield dynamics with these
experimental characteristics.

\section{QMC simulation results and phase diagram}

{}From the QMC simulations, we have compiled a ``phase
diagram'' for the RC, which is a map of the various regions in
parameter space which generate dynamics with all the key experimental
characteristics.  This phase diagram is shown in Fig.~1, with the
proper values of \K12 correlated with the energy of the P$^+$B$^-$ state
\E2.
On the phase diagram, we have also indicated the values of \K12
predicted by conventional nonadiabatic theory to give a
room-temperature ET rate of 3~ps (without demanding that they
also conform to the inverse temperature dependence).

{}From the phase diagram,
there are clearly two distinct regions which yield charge separation
dynamics with the correct experimental characteristics.
The first is characterized by a P$^+$B$^-$ state energy \E2 lower
than about $-600$~\cm\, and down to about $-1300$~\cm.  Within this region,
the dynamics largely agrees with predictions from
conventional nonadiabatic theory for the sequential mechanism.
Simulation results for the transient populations on the three electronic
states are shown in Fig.~2 at two temperatures for \E2 = $-666$~\cm\,
and $-1333$~\cm.  The 1 $\to$ 3 transfer is efficient in both
cases, with \P2 remaining below 15\% throughout.
For \E2 = $-666$~\cm, the charge separation rate increases from 3.0~ps
at 298~K to 2.0~ps at 140~K.
This should be contrasted with \E2 = $-1333$~\cm, for which the
rate increases from 3.3~ps at 298~K to only 2.7~ps at 140~K.
Although the ET rate still increases with lower
temperature for \E2 = $-1333$~\cm, this temperature-dependence
is somewhat weaker than that observed in experiments.  For
\E2 much below $-1333$~\cm, the temperature dependence is not strong
enough to be consistent with experiments.  Therefore, in the phase
diagram, the sequential region terminates around \E2 = $-1333$~\cm.

\vbox{
\begin{figure}
\epsfig{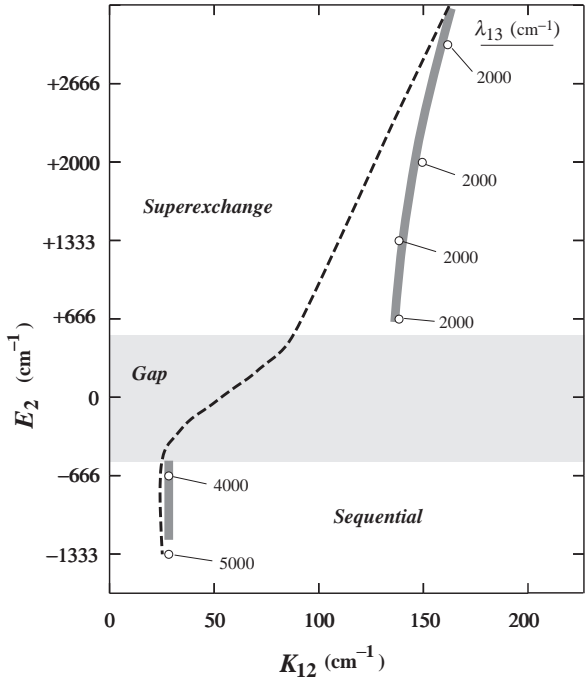}{4.00in}{0.00in}{40 0 444 526}
\caption[]{\label{fig1} Phase diagram of the RC
compiled from QMC simulation data, correlating
$K_{12}$ with $E_2$ (thick lines).
The proper reorganization energies
are indicated on the phase diagram
for those parameter sets
that yield dynamics consistent with experiments (open circles).
Nonadiabatic predictions for the \K12
yielding a room-temperature rate of 3~ps
are also shown for comparison (dashed curve).}
\end{figure}
}

The second region is characterized by a P$^+$B$^-$ state energy \E2
higher than $+666$~\cm\, (we have performed simulations for
\E2 up to $+3000$~\cm).  For \E2 = $+666$~\cm\, and $+1333$~\cm,
conventional nonadiabatic superexchange theory does {\em not}\, give
reliable predictions for the dynamics.  The transfer mechanism in
this region is predominantly superexchange, although the precise rates
are not well described by nonadiabatic theory.
As the energy of the P$^+$B$^-$ state moves up to \E2 = $+2000$~\cm\,
and beyond, the dynamics becomes increasingly nonadiabatic.
We mention in passing that in this limit the dynamical sign problem is quite
severe, and one needs to sample extremely long Monte Carlo
trajectories rendering QMC simulations for \E2 much higher than $+3000$~\cm\,
very costly. Nevertheless, our present data show that
for sufficiently high-lying P$^+$B$^-$ state, the conventional
golden rule superexchange formula becomes increasingly accurate,
making QMC simulations less valuable for \E2 $>$ $+3000$~\cm.
For \E2 $< +2000$~\cm, however,
nonadiabatic theory seems to severely underestimate the magnitude of
the electronic couplings  required to achieve superexchange
rates consistent with experiments.

\vbox{
\begin{figure}
\epsfigrot{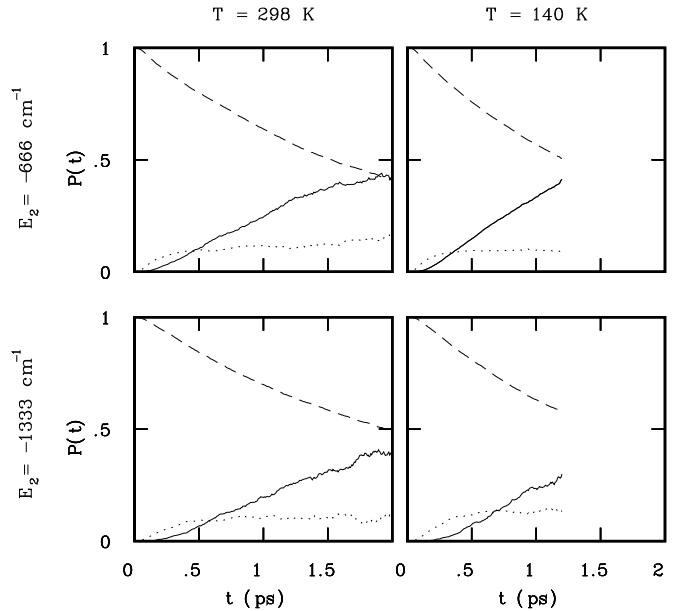}{3.50in}{0.30in}{120 -600 540 -200}
\caption[]{\label{fig2}  QMC data for the ET dynamics with low-lying
P$^+$B$^-$ state and $K_{12}=28$~\cm\,
for $E_2=-666$~\cm\, (\L13$=4000$~\cm)
and $E_2=-1333$~\cm\, (\L13$=5300$~\cm). The dashed (dotted, solid)
curve gives the occupation probability on state 1 (2, 3).}
\end{figure}
}

Simulation results for the transient populations on the three
electronic states are shown in Fig.~3 at two temperatures for
\E2 = $+666$~\cm, $+1333$~\cm\, and $+2000$~\cm.  For all three
energies, the charge separation rates at 298~K are about 3~ps.
These rates increase to about 2.2 to 2.5~ps at 140~K.  As was pointed
out in our previous work \cite{JPC}, the dynamics in this regime is
not simply monoexponential.  There is a fast short-time component
for the first $\approx 0.5$~ps, followed by a slower component.  The
fast component always obeys the experimentally observed inverse
temperature dependence, but the rate and the proportion of the slower
component vary nonmonotonically with temperature, as measured
in many mutant RCs \cite{mutants}. The physical reason
for this phenomenon has been given by Gehlen et al.\cite{david}, and
our simulations provide direct numerical support for their arguments.

The phase diagram in Fig.~1 also reveals another interesting aspect of the
superexchange regime: the RC seems to be able to
operate under a wide range of P$^+$B$^-$ state energies
with the same conditions, i.e.,
with approximately the same electronic coupling \K12 (about
140 to 150~\cm) and the same reorganization energy \L13 (about 2000~\cm).
This means that within the superexchange regime, the system is
robust against variation in the P$^+$B$^-$ state energy.
The sequential region is not as robust.
Although approximately the same value of \K12 (about 28~\cm) would work
for \E2 between about $-1333$~\cm\, and $-666$~\cm, the optimal value of the
reorganization energy \L13 changes from 4000~\cm\, at \E2 = $-666$~\cm\,
to 5300~\cm\, at \E2 = $-1333$~\cm.

\begin{figure}
\epsfigrot{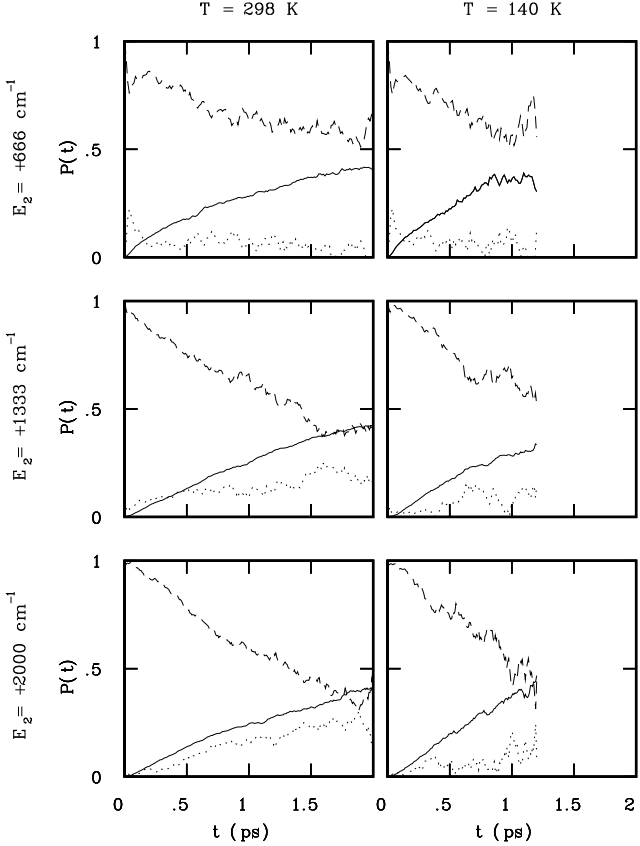}{4.50in}{0.30in}{-20 -600 540 -200}
\caption[]{\label{fig3} QMC data for high-lying P$^+$B$^-$ state with
\L13$=2000$~\cm. The couplings are $K_{12}=140$~\cm\, for
$E_2=+666,+1333$~\cm, and $K_{12}=150$~\cm\, for
$E_2=+2000$~\cm.}
\end{figure}

Finally, it is important to stress that the RC does {\em not}\, seem to
operate via a combined sequential/superexchange mechanism as
proposed in Refs.\cite{bixon91,chan}. Our simulations indicate
clearly that it is either a pure sequential or a pure superexchange
transfer.  This is indicated by the grey region in Fig.~1, which we
refer to as the ``gap'' region.
We emphasize that this does not imply that in this region it is impossible
to achieve a fast ET rate.  On the contrary, it is possible to achieve
a 3~ps transfer at room temperature with only minor population accumulation
on the P$^+$B$^-$ state.  However, we have not found any parameter
set in this region that can simultaneously satisfy all the experimentally
observed characteristics {\em both} at high and low temperatures
\cite{JPC}.

\section{Noninteracting-cluster approximation}

By invoking a simple and reasonable approximation for the
formally exact path-integral expressions,
we have previously derived a set of nonlocal master equations
for the time-dependent occupation probabilities \cite{nica}.
Our ``noninteracting-cluster approximation'' (NICA)
expresses the time-dependent transition
 rates between the various sites in terms
of cluster functions $\Gamma_{ij}$.
These cluster functions
give the amplitude of all paths going from diagonal state $i$ to
diagonal state $j$ of the reduced density matrix without touching
the diagonal in between. Since each hop on the
$3\times 3$ lattice representing the possible states of the
reduced density matrix gives a factor $\pm iK_{ij}$, the
rates can be expressed as a power series in the electronic
couplings.

By confining this expansion to the lowest-order terms and
taking the long-time limit (which makes
the master equations local in time), one
obtains the nonadiabatic rate expressions referred to in Sec.~III.
The rates $\Gamma_{12}$ and $\Gamma_{23}$
(plus backward rates) describing a sequential transfer are
just golden rule rates. Furthermore, the lowest-order cluster
$\Gamma_{13}^{(4)}\sim K_{12}^2 K_{23}^2$
gives the nonadiabatic superexchange rate.
There is only one pathway contributing to this fourth-order
cluster (plus the complex conjugate),
namely the one going along the edge of the $3\times 3$ lattice.
The resulting rate formula can be further simplified
in the classical limit by taking a short-time approximation
for the bath kernel, or for a high-lying state where
one can derive the conventional golden rule rate for the emerging two-state
system spanned by states 1 and 3 \cite{nica}.
These golden rule rates are consistent with the
QMC results
for very high P$^+$B$^-$ state energy ($E_2 \gapx +2000$~\cm)
as well as for low-lying P$^+$B$^-$ states ($E_2 \lapx -666$~\cm).
But in the intermediate region $+666$~\cm \lapx \E2 \lapx $+2000$~\cm,
nonadiabatic theory breaks down (see Fig.~1).

We will now study the origin of the
apparent discrepancy between  nonadiabatic
theory and the QMC results in this intermediate region.
To that purpose, we have computed the next-order corrections
to the superexchange cluster, $\Gamma_{13}^{(6)}$.
Here one has to consider the two pathways
with six hops which go from diagonal state 1 to state 3 without hitting
the diagonal.  In the classical limit, we obtain
for the activationless situation considered in the simulations
 ($E_3 =-\lambda_{13}$)
\begin{equation}
\frac{\Gamma_{13}^{(6)}}{\Gamma_{13}^{(4)}} =
- \frac{K_{12}^2 + K_{23}^2}{4 (\delta E)^2} \left(
1 - \frac{2\beta F^*}{1+\exp(\beta F^*)} \right) \;,
\end{equation}
where $\delta E= E_2 + \lambda_{13}/4$ and
$F^*=(\delta E)^2/\lambda_{13}$. For a high-lying
intermediate state, the bracket gives unity because
$\beta F^* \gg 1$. Therefore the sixth-order correction
decreases the nonadiabatic estimate by
$\approx 20\%$ for $E_2=+666$~\cm. With the
short-time bath kernel, it would actually be possible to sum the whole
power series and thus to obtain
the full classical NICA superexchange rate
without stopping at sixth order.
However,
we have not done so because
this short-time approximation is inappropriate
for high-order diagrams. In any case, it is expected that these
higher-order contributions lead to a {\em reduction} of the
nonadiabatic superexchange rate.
Moreover, we also expect that adiabatic corrections beyond
NICA can lead to substantial renormalizations
for large electronic couplings.

To conclude, we have given a path-integral analysis
of the primary electron transfer step in bacterial
photosynthesis. The mechanism appears to be either a sequential
or a superexchange transfer but not a combined one.
Whereas conventional nonadiabatic theory appears to
be accurate for both the sequential and the ultimate
superexchange regime, adiabatic corrections
reduce the superexchange rate substantially for
intermediate energies of the P$^+$B$^-$ state.

We wish to thank David Chandler and Uli Weiss for illuminating
discussions.
This work was supported in part by
the National Science Foundation (CHE-9216221)
and the NSF Young Investigator Awards Program (CHE-9257094),
the Camille and Henry Dreyfus Foundation under the Camille
Teacher-Scholar Award Program,
and
the Sloan Foundation through a Alfred P. Sloan Fellowship.
Computational resources from IBM are acknowledged.

\end{document}